\documentclass[preprint,12pt]{elsarticle}

\usepackage{amssymb}
\usepackage{amsmath} 
\usepackage{url}
\usepackage[multiple]{footmisc}

\journal{Computer Physics Communications}

\begin{document}
\newcommand{\asymop}{\mathcal{A}}
\newcommand{\densmat}[3]{\rho^{\mathrm{#1}}_{#2 #3}}
\newcommand{\densmatmata}[2]{\boldsymbol{\rho}^{\mathrm{#1}}_{#2}}
\newcommand{\densmatmatb}[2]{\boldsymbol{\varrho}^{\mathrm{#1}}_{#2}}
\newcommand{\densmatb}[2]{\varrho^{\mathrm{#1}}_{#2}}

\newcommand{\psik}[2]{\psi_{#1}^{(N+1)\mathrm{#2}}}
\newcommand{\psitg}[2]{\Phi_{#1}^{(N)\mathrm{#2}}}
\newcommand{\phicsflsq}[2]{\phi_{#1}^{(N+1)\mathrm{#2}}}
\newcommand{\phicsftg}[2]{\phi_{#1}^{(N)\mathrm{#2}}}
\newcommand{\csflsq}[2]{\chi_{#1}^{(N+1)\mathrm{#2}}}
\newcommand{\csftg}[2]{\chi_{#1}^{(N)\mathrm{#2}}}

\newcommand{\Psiscat}[2]{\Psi_{#1}^{(\pm)\mathrm{#2}}}
\newcommand{\Psibound}[2]{\Phi_{#1}^{(N+1)\mathrm{#2}}}

\newcommand{\ctorb}[1]{\eta_{#1}}
\newcommand{\dyson}[3]{D^{\mathrm{#1}}_{#2 #3}}

\newcommand{\xtot}{x_1,...,x_{N},x_{N+1}}
\newcommand{\xtarg}{x_1,...,x_{N}}
\newcommand{\dx}{dx_1...dx_N}

\newcommand{\mat}[1]{\mathbf{#1}}
\begin{frontmatter}



\title{CDENPROP: Transition matrix elements involving continuum states}


\author{Alex G. Harvey}
\ead{harvey@mbi-berlin.de}
\author{Danilo S. Brambila}
\author{Felipe Morales}
\author{Olga Smirnova}
\address{Max-Born-Institut, Max-Born-Strasse 2A, D-12489 Berlin, Germany}

\begin{abstract}
Transition matrix elements between electronic states where one electron can 
be in the continuum are required for a wide range of applications of the 
molecular R-matrix method.
These include photoionization, photorecombination and photodetachment; 
electron-molecule scattering and photon-induced processes in the presence 
of an external D.C. field, and time-dependent R-matrix approaches to study the 
effect of the exposure of molecules to strong laser fields.
We present a new algorithm, implemented as a module (CDENPROP) in the UKRmol 
electron-molecule scattering code suite.
\end{abstract}


\end{frontmatter}


\section{Introduction}

Bound-continuum and continuum-continuum transition dipoles are necessary for  
photoionization, photorecombination, scattering and photon-induced processes
in a D.C. field, and the time-dependent implementation of the R-matrix method.
However, the UKRmol \cite{carr12} codes use a basis set expansion of the 
$(N+1)$-electron scattering wavefunction (and the $(N+1)$-electron bound states) 
which has a subset of basis functions that are not Slater determinants but 
linear combinations of them \cite{tennyson96}, called the contracted basis. 
This leads to much smaller Hamiltonians and hence significant speed up in 
Hamiltonian diagonalisation but means that one can no longer use Slater's 
rules (see, for example, \cite{mcweeny69}) in the standard method of 
evaluating matrix elements as used in the existing UKRmol bound state 
transition moments code, DENPROP. This paper describes a new code, CDENPROP, 
that works with the contracted basis to produce transition matrix elements 
between members of the contracted basis set, our method also allows us to 
easily construct Dyson orbitals.

\section{Theory}
\subsection{The R-matrix method}

We will make no attempt at a comprehensive description of the R-matrix method,
as this is a topic that has been covered in depth elsewhere 
\cite{tennyson10,burke11}.
Instead, we will give a brief description of the method and then focus only on 
those parts that are necessary to elucidate the description of the new code.

The key feature of the R-matrix method is the division of space into separate 
regions in which different approximations may be made.
The usual division is into an inner, outer and asymptotic region.
The boundary between the inner and outer region is defined such as to fully 
contain the bound wavefunctions of the molecule. 
In the inner region the electron-electron effects, exchange, polarisation and 
correlation, are fully accounted for in a manner analogous to a quantum 
chemistry calculation, and a flexible basis is constructed to 
describe the scattering or bound wavefunction in the inner region.
The R-matrix is defined on the boundary and relates the radial part of the 
continuum electron wavefunction to its derivative.
In the outer region exchange is neglected and the continuum electron moves in 
the long range multipole potential of the molecule.
The R-matrix is propagated through the outer region and matched to known 
asymptotic solutions, allowing for the calculation of observables and giving 
(with some extra work \cite{harvey13a}) the expansion coefficients of the full 
multi-electron wavefunction in terms of the inner region basis.  

\subsection{The inner region basis}

The inner region wave functions, $\psik{k}{}$, are represented by a close 
coupling expansion as follows, 
\begin{eqnarray}\label{eqclosecoup}
\psik{k}{}&=&\asymop \sum_{i m_i} a_{k i m_{\gamma_i}}\psitg{i}{}(\xtarg)\ctorb{m_{\gamma_i}}(x_{N+1}) \nonumber \\ 
          &+& \sum_{p} b_{kp} \csflsq{p}{}(\xtot),
\end{eqnarray}
where $\psitg{i}{}$ is a $N$ electronic target state, generally produced by some 
CI procedure, $\ctorb{m_{\gamma_i}}$ is a continuum orbital, the subscript of
the continuum orbital index  $m_{\gamma_i}$ labels the symmetry of the continuum 
orbital and is determined by the symmetry of the target state and the overall 
symmetry. 
$\asymop$ is the antisymmetrisation operator. 
In the second summation $\csflsq{p}{}$ is a $N+1$ electron $L^2$ 
configuration constructed by placing the continuum electron into 
the bound orbital set, this is required to recover short range 
correlation/polarisation lost due to the prior orthogonalisation
of bound and continuum orbitals. 
$a_{k i m_{\gamma_i}}$ and $b_{pk}$ are variationally determined 
coefficients. 
We note that all quantities in eq. (\ref{eqclosecoup}) are real.
Outer region propagation and matching to asymptotics allows both continuum and
bound state coefficients to be determined:
\begin{eqnarray}\label{eqcontinuumstate}
\Psiscat{i}{}=\sum_k A^{(\pm)}_{ik}\psik{k}{}
\end{eqnarray}
\begin{eqnarray}\label{eqboundstate}
\Psibound{i}{}=\sum_k B_{ik}\psik{k}{},
\end{eqnarray}
where $\Psiscat{i}{}$ is a continuum state with outgoing or incoming boundary 
conditions in channel $i$ and $\Psibound{i}{}$ is a bound state formed by 
considering all scattering channels to be closed \cite{sarpal91}.

\subsection{The density matrix}
\label{}

All one electron properties can be calculated from the one-electron transition
density matrix, 
\begin{eqnarray}\label{eqdensity1}
\densmat{N+1}{k}{l}(x_{N+1})=(N+1)\int \psik{k}{}(\xtot)\psik{l}{}(\xtot) \dx,
\end{eqnarray}
and constructing it is half the battle. 
In this section we see how to take advantage of the structure of the inner 
region basis to simplify its calculation. 
Inserting eq. (\ref{eqclosecoup}) into eq. (\ref{eqdensity1}) we get

\begin{eqnarray}\label{eqdensity2}
\densmat{N+1}{k}{l}&=& (N+1)\int  \sum_{i m_{\gamma_i}}\sum_{j n_{\gamma_j}} a_{k i m_{\gamma_i}}a_{l j n_{\gamma_j}}\asymop[\psitg{i}{}\ctorb{m_{\gamma_i}}] \asymop[\psitg{j}{}\ctorb{n_{\gamma_j}}] \dx \nonumber \\
                   &+& (N+1)\int  \sum_{i m_{\gamma_i}}\sum_{q} a_{k i m_{\gamma_i}}b_{ql}\asymop[\psitg{i}{}\ctorb{m_{\gamma_i}}]  \csflsq{q}{} \dx \nonumber \\
                   &+& (N+1)\int  \sum_{p}\sum_{j n_{\gamma_j}} b_{pk}  a_{l j n_{\gamma_j}}\csflsq{p}{})\asymop[\psitg{j}{}\ctorb{n_{\gamma_j}}] \dx \nonumber \\
                   &+& (N+1)\int \sum_{p}\sum_{q} b_{pk} b_{ql} \csflsq{p}{}  \csflsq{q}{} \dx,
\end{eqnarray}
which we can write in matrix form as
\begin{eqnarray}\label{eqdensity2.1}
\densmatmata{N+1}{}{}=
  \begin{bmatrix}
    \mat{a} & \mat{b} \\
  \end{bmatrix}
  \begin{bmatrix}
    \densmatmatb{C-C}{}{} & \densmatmatb{C-L^2}{}{} \\
    \densmatmatb{L^2-C}{}{} & \densmatmatb{L^2-L^2}{}{} \\
  \end{bmatrix}
  \begin{bmatrix}
    \mat{a}  \\
    \mat{b}  \\
  \end{bmatrix}.
\end{eqnarray}
We look first at the diagonal blocks of the matrix $\densmatmatb{}{}{}$.
The $\mathrm{L^2-L^2}$ block has elements, 
\begin{eqnarray}
\densmatb{L^2-L^2}{pq}{}=(N+1)\int \csflsq{p}{}  \csflsq{q}{} \dx,
\end{eqnarray}
which have $\mathrm{L^2}$ functions for both initial and final state, and can be evaluated 
using Slater's rules.
The $\mathrm{C-C}$ block, 
\begin{eqnarray}
\densmatb{C-C}{im_{\gamma_i} j n_{\gamma_j}}{}= (N+1)\int \asymop[\psitg{i}{}\ctorb{m_{\gamma_i}}] \asymop[\psitg{j}{}\ctorb{n_{\gamma_j}}] \dx,
\end{eqnarray}
in which both initial and final states contain continuum orbitals, can be 
reduced, using the  orthonormality of bound and continuum orbitals, and of 
bound states, to 
\begin{eqnarray}
\densmatb{C-C}{im_{\gamma_i} j n_{\gamma_j}}{}=\delta_{ij}\ctorb{m_{\gamma_i}} \ctorb{n_{\gamma_j}} +\delta_{m_{\gamma_i} n_{\gamma_j}}\densmat{N}{i}{j}
\end{eqnarray}
where $\densmat{N}{i}{j}$ is the transition density matrix for the target 
molecule.

Now considering the off diagonal blocks,
\begin{eqnarray}
\densmatb{C-L^2}{im_{\gamma_i}q}{} &=& (N+1)\int \asymop[\psitg{i}{}\ctorb{m_{\gamma_i}}]  \csflsq{q}{} \dx \nonumber \\
\densmatb{L^2-C}{pj n_{\gamma_j}}{} &=& (N+1)\int  \csflsq{p}{})\asymop[\psitg{j}{}\ctorb{n_{\gamma_j}}] \dx, 
\end{eqnarray}
which become
\begin{eqnarray}
\densmatb{C-L^2}{im_{\gamma_i}q}{} &=& \ctorb{m_{\gamma_i}} \dyson{\chi}{i}{q} \nonumber \\
\densmatb{L^2-C}{pj n_{\gamma_j}}{} &=& \dyson{\chi}{p}{j}\ctorb{n_{\gamma_j}},
\end{eqnarray}
where $\dyson{\chi}{i}{q} = \sqrt{N+1}\int \psitg{i}{}\csflsq{q}{}\dx$.
To evaluate $\dyson{\chi}{i}{q}$ we need to first expand the target states in 
their CSF basis, $\psitg{i}{}=\sum_r c_{ir}\csftg{r}{}$.
\begin{eqnarray}\label{eqdyson1}
\dyson{\chi}{i}{q} &=& \sqrt{N+1}\int \sum_{r} c_{ir}\csftg{r}{}\csflsq{q}{} \dx \nonumber \\
                   &=& \sum_{r} c_{ir} \dyson{\chi\chi}{r}{q}, 
\end{eqnarray}
where $\dyson{\chi\chi}{r}{q} =\sqrt{N+1}\int \csftg{r}{}\csflsq{q}{}\dx$. 
This is straightforward to calculate in principle by checking if 
the set of target orbitals in $\csftg{r}{}$ is a subset of the orbitals in 
$\csflsq{q}{}$.
In practice however, due to the manner in which the contracted basis is 
initially created, it is simpler to use Slater's rules on the integral 
$(N+1)\int \csftg{r}{}\ctorb{m_{\gamma_i}}\csflsq{q}{}\dx=\ctorb{m_{\gamma_i}}\dyson{\chi\chi}{r}{q}$

\subsection{Transition moments}

Once $\densmatmatb{}{}{}$ has been constructed we can construct the moments
matrix, $\mathcal{M}$, as follows,
\begin{eqnarray}
\mathcal{M}=\int \mu \densmatmatb{}{}{} dx,
\end{eqnarray}
where $\mu$ is the transition moment operator (dipole, quadrapole etc)
The final step in calculating the transition moments between the inner region 
basis functions, $\mathbf{M}$, is then simply to multiply in the coefficient
matrices.
\begin{eqnarray}\label{eqdensity5}
\mathbf{M}=
  \begin{bmatrix}
    \mat{a} & \mat{b} \\
  \end{bmatrix}
    \bf{\mathcal{M}}
  \begin{bmatrix}
    \mat{a}  \\
    \mat{b}  \\
  \end{bmatrix}.
\end{eqnarray}

\subsection{Dyson orbitals}

Dyson orbitals may be obtained in a straight forward manner, taking the case 
that the initial state is bound,  
\begin{eqnarray}
\mat{\ctorb}\mat{D}=
  \begin{bmatrix}
    \densmatmatb{C-C}{}{} & \densmatmatb{C-L^2}{}{} \\
  \end{bmatrix}
  \begin{bmatrix}
    \mat{a}  \\
    \mat{b}  \\
  \end{bmatrix}
   \mat{B},
\end{eqnarray}
and discarding the continuum orbital, $\mat{\ctorb}$, that comes from the final 
state, elements of the matrix $\mat{D}$ are the Dyson orbitals,
\begin{eqnarray}
\dyson{}{i}{j} = \sqrt{N+1}\int \psitg{i}{}\Psibound{j}{}\dx.
\end{eqnarray}
\section{The code}
\begin{figure}
\includegraphics[width=145mm]{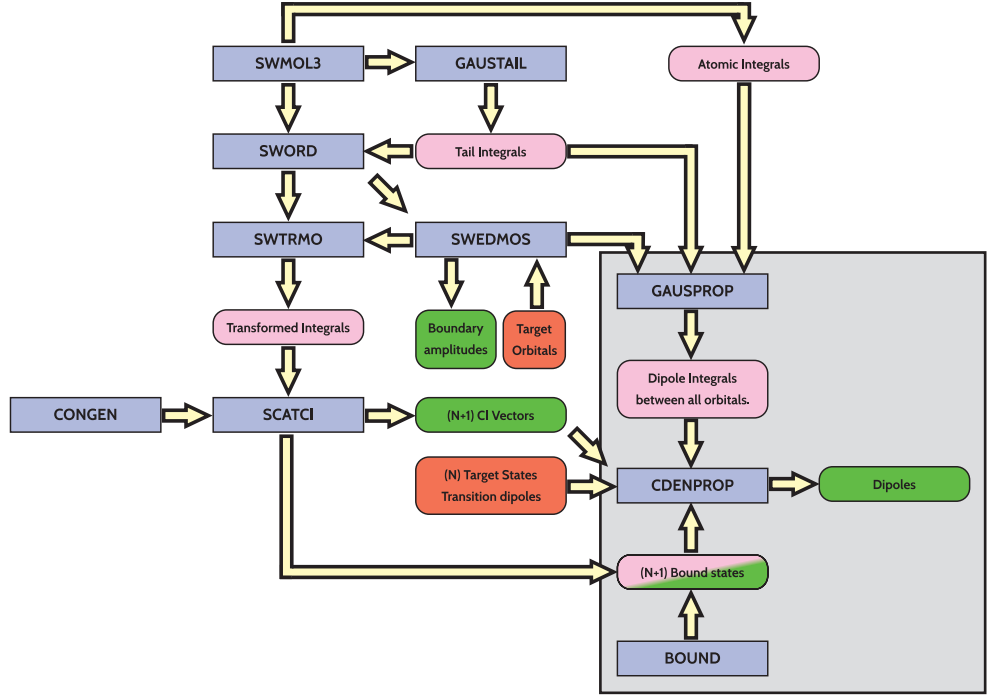}%
 \caption{Schematic of the UKRmol inner region codes with the addition of 
 CDENPROP. The shaded rectangle encloses the routines that are required to
 produce transition matrix elements. Red boxes indicate input from the target
 calculation, pink boxes input/output for inner region use, and green boxes
 represent input/output from/to the outer region. The $(N+1)$ Bound states box 
 is both green and pink indicating that these states may be constructed either
 using outer region, or inner regions codes.} 
\label{fig_inner_region_plus_cdenprop}
\end{figure}
The UKRmol code suite\footnote{\url{http://ccpforge.cse.rl.ac.uk/gf/project/ukrmol-in}}\footnote{\url{http://ccpforge.cse.rl.ac.uk/gf/project/ukrmol-out}} 
(including CDENPROP) is a project registered on CCPForge\footnote{\url{http://ccpforge.cse.rl.ac.uk/gf}}, 
providing an online repository which allows for collaborative work using the 
Subversion version control system. 

A calculation is performed in 3 stages, first the target electronic 
states are constructed and transition moments between them, then the inner 
region Hamiltonian is constructed and solved, finally the R-matrix is 
constructed, propagated in the outer region and matched to asymptotics. 
CDENPROP resides primarily in the inner region stage,  
fig. (\ref{fig_inner_region_plus_cdenprop}) shows a schematic of the various 
inner region codes with the most important inputs and outputs. The shaded 
rectangle indicates steps required to produce inner region dipoles, including
the new module CDENPROP that implements the theory of the preceding section.
CDENPROP takes as input, target states and transition dipoles from the target 
calculation, dipole integrals between both bound and continuum orbitals 
restricted to the inner region, inner regions wavefunctions (the $(N+1)$ CI 
vectors in fig. (\ref{fig_inner_region_plus_cdenprop})) and, optionally, bound 
states produced by the outer region code BOUND. 
It outputs the inner region dipoles and Dyson orbitals. We note that CDENPROP 
is also capable of calculating transition dipoles between the target states.

CDENPROP borrows routines from the existing UKRmol code, DENPROP, for the 
application of Slater's rules.
Aside from the construction of the density matrix, there is another key 
difference between the two codes.
DENPROP constructs the density matrix for each state pair, reducing the 
density matrix in symbolic form to a small set of orbitals pairs and 
corresponding coefficients, and then picks up 
the relevant dipole integrals, a procedure that is memory efficient but 
scales like $\mathcal{O}(n^4)$ with the basis set size, $n$, this is ameliorated by the 
sparseness of the density matrix when each basis function is a single 
Slater determinant, and by the fact that, generally, during a target run, 
only a small subset of the target states are required. 
Neither of these conditions hold in the inner region, where the contracted 
basis leads to the $\mathrm{C-L^2}$ blocks being non-sparse and where dipoles 
between all the inner region wavefunctions may be required. The procedure 
DENPROP uses rapidly becomes unfeasible as the basis set size increases.
The technique outlined in the previous section picks up the dipole integrals 
first, and then multiplies in the state coefficients, giving a scaling of 
$\mathcal{O}(n^3)$ and only requiring a single pickup of the dipole integrals, but 
paying a penalty in increased memory requirements. We use sparse matrix 
routines throughout the code where appropriate.
It is important to note that the memory requirements are of the same order as the 
Hamiltonian construction and diagonalisation code SCATCI \cite{tennyson96} 
and so memory issues tend to show up, and are addressed, at this stage, 
prior to reaching CDENPROP.

\section{Conclusions}

We have described our newly developed technique for calculating the inner 
region transition matrix elements needed in a wide range of applications. 
The technique is implemented in the code module, CDENPROP, and takes 
advantage of the structure of the inner region basis to efficiently calculate 
the transition matrix elements.
Finally, we note that we have successfully applied the new code in studies 
on angular resolved photoionization from aligned CO$_2$ 
\cite{rouzee13,harvey13a}, He photoionization and electron-He$^+$ scattering 
in a weak D.C. external field (Brambila et al. in prep.) and to describe the 
recombination step in high harmonic generation experiments (Harvey et al. in 
prep.). 
The new code also forms an integral part of the ongoing development of the 
molecular time-dependent R-matrix approach.

\section*{Acknowledgements}
The authors would like to acknowledge useful discussions with Michal Tarana
and Jonathan Tennyson.
We acknowledge the support of the Einstein foundation project A-211-55 Attosecond 
Electron Dynamics.

\bibliographystyle{elsarticle-num}
\bibliography{cdenprop,sub-prep-arxiv}







\end{document}